\documentclass{article}

\usepackage{graphicx}
\usepackage{amsmath,amsfonts,amssymb}
\usepackage{euler}
\usepackage{epstopdf}
\usepackage{hyperref}
\hypersetup{
    colorlinks=true,
    citecolor=blue,
    filecolor=blue,
    linkcolor=red,
    urlcolor=blue
}
\usepackage[caption=false,font=footnotesize]{subfig}
\usepackage{epsfig,epsf,psfrag}

\newcommand{\bqn}{\begin{eqnarray}}
\newcommand{\eqn}{\end{eqnarray}}
\newcommand{\bq}{\begin{eqnarray*}}
\newcommand{\eq}{\end{eqnarray*}}

\title{Mapping Brain-Behavior Correlations in Autism Using Heat Kernel Smoothing\footnote{Originally published in Quantitative Bio-Science 30(2), 75-83 (2011).}}

\author{
Moo K. Chung\textsuperscript{1},
Kim M. Dalton\textsuperscript{2},
Daniel J. Kelley\textsuperscript{2},\\
Richard J. Davidson\textsuperscript{3}
}

\date{
\small
\textsuperscript{1}Department of Biostatistics and Medical Informatics, University of Wisconsin–Madison, Madison, WI 53705, USA\\
\textsuperscript{2}Waisman Center, University of Wisconsin–Madison, Madison, WI 53705, USA\\
\textsuperscript{3}Department of Psychology and Psychiatry, University of Wisconsin–Madison, Madison, WI 53705, USA
}

\begin{document}

\maketitle

\begin{abstract}
This paper presents a streamlined image analysis framework for
correlating behavioral measures to anatomical measures on the cortex
and detecting the regions of abnormal brain-behavior correlates. We
correlated a facial emotion discrimination task score and its
response time to cortical thickness measurements in a group of high
functioning autistic subjects. Many previous correlation studies in
brain imaging neglect to account for unwanted age effect and other
variables and the subsequent statistical parametric maps may report
spurious results. We demonstrate that the partial correlation
mapping strategy proposed here can remove the effect of age and
global cortical area difference effectively while localizing the
regions of high correlation difference. The advantage of the
proposed correlation mapping strategy over the general linear model
framework is that we can directly visualize more intuitive
correlation measures across the cortex in each group.
\end{abstract}

%\begin{keyword}
%Cortical thickness \sep heat kernel \sep diffusion smoothing \sep partial correlation \sep autism
%\end{keyword}

\section{Introduction}

Pearson's product-moment correlation (Fisher, 1915), in short simple
correlation, has widely been used as a simple index for measuring
dependency and the linear relationship between two variables. In
human brain mapping research, it has been mainly used to map out
functional or anatomical connectivity (Friston et al., 1993; Horwitz
et al., 1996; Friston et al., 1996; Cao and Worsley, 1998; Worsley
et al., 2005). In this framework, correlations between pairs of
voxels are computed and thresholded via the random field theory to
reveal the statistically significant regions of connectivity by
testing the existence of correlation $\rho$ on the template cortex
\bqn
H_0: \rho(p) = 0 \quad \text{for all } p \in \partial \Omega \quad \text{vs.} \quad
H_1: \rho(p) \neq 0 \quad \text{for some } p \in \partial \Omega
\label{eq:multi}
\eqn

In a different setting, Thompson et al. (2001) used the correlation
between genetic factors and the amount of gray matter on the cortex
via a linear model in mapping out the regions of genetic influence.
Our use of correlation is somewhat similar to Thompson et al. (2001)
in that we correlate anatomical index to non-anatomical index on the
cortex. In this study, we map out the dependency of behavioral
measures to an anatomical measure spatially over the cortex and
localize the regions of abnormal correlation difference between
groups. To remove unwanted covariates like age and total brain size
difference, we introduce the concept of {\em partial correlation
coefficient}, in short partial correlation. Chung et al. (2004;
2005) has already demonstrated the need for removing the effect of
age and global brain size difference in morphometric analyses so it
is crucial to use partial correlation rather than the usual simple
correlation in our study. Although our correlation mapping strategy
can be formulated in terms of a general linear model (GLM) as in the
case of Thompson et al. (2001), our unified approach will provide a
more intuitive alternative that is visually comprehensive.

As an application, we applied our method in characterizing abnormal
brain-behavior correlation in autism. We correlated two behavioral
measures with the anatomical measure, {\em cortical thickness}. The
cortical thickness measures the thickness of the gray matter shell
bounded by the both outer and inner cortical surfaces (MacDonald et
al., 2000; Chung et al., 2003; Chung et al., 2005). The first
behavioral measure is the emotional face recognition task score. The
task score counts the number of correct responses when judging
whether a subject is viewing an emotional (happy, fear and anger) or
neutral face (Dalton et al., 2005). The second behavioral measure is
the time required to produce a response. The response time is
measured in ms. Each behaviorial measure was correlated with the
cortical thickness measure at each point on the cortex for the both
autistic and control groups, and a statistical test was performed to
determine the regions of differing correlation pattern between
groups. This study is a continuation of the series of multifaceted
studies in the Waisman laboratory for brain imaging and behavior
characterizing the autistic, structural, functional, and behavioral
phenotypes (Chung et al., 2004; Chung et al.,2005; Dalton et al.,
2005).

\section{Prelimary}
Let $Y=(Y_1,Y_2)$ be two variables of interests and
$X=(X_1,\cdots,X_p)$ be a row vector of variables that should be
removed in a data analysis. For instance, we may let $Y_1$ be the
cortical thickness, $Y_2$ be the response time, and $X_1$ and $X_2$
be the age and total surface area respectively. The covariance
matrix of $(Y,X)'$ is denoted by
\bqn \mathbb{V} (Y,X)' = \left(%
\begin{array}{cc}
  \Sigma_{YY} & \Sigma_{YX} \\
  \Sigma_{XY} & \Sigma_{XX} \\
\end{array}\right) \label{eq:cov} \eqn Note $\Sigma_{XY}$ is the
cross-covariance matrix of $X$ and $Y$. $\Sigma_{YX}$ and
$\Sigma_{XX}$ are defined similarly. Then the partial covariance of
$Y$ given $X$ is
$$\Sigma_{YY} - \Sigma_{YX}\Sigma_{XX}^{-1}\Sigma_{XY} =(\sigma_{ij}).$$
The {\em partial correlation} $\rho_{Y_i,Y_j|X}$ is the correlation
between variables $Y_i$ and $Y_j$ while removing the effect of
variables $X$ and it is defined as

$$\rho_{Y_i,Y_j|X} = \frac{\sigma_{ij}}{\sqrt{\sigma_{ii}\sigma_{jj}}}.$$
The {\em conditional} notation $|$ is used in defining the partial
correlation since the partial correlation is equivalent to {\em
conditional correlation} if $\mathbb{E} (Y|X) = a + B X$ for some
vector $a$ and matrix $B$, which is true under the normality of
data. This is the formulation we used to compute the partial
correlation. If vector $X$ consists of a single measurement, i.e.
$X=X_1$, the partial correlation can be computed from the simple
correlation via

$$\rho_{Y_1,Y_2|X} = \frac{\rho_{Y_1,Y_2} - \rho_{Y_1,X}\rho_{Y_2,X}}
{\sqrt{(1-\rho_{Y_1,X}^2)(1-\rho_{Y_2,X}^2)}}.$$ The {\em sample
partial correlation} $r_{Y_1,Y_2|x}$ is defined similarly by
replacing the covariance with the sample covariance in
(\ref{eq:cov}). 

If we let $\rho_k$ be the partial correlation for group $k$ ( autism
$=1$, control $=2$), for each fixed $p \in
\partial \Omega$, one may test
\bqn H_0^A: \rho_k(p) = 0 \mbox{ vs. } H_1^A:\rho_k(p) \neq
0\label{eq:onesample}.\eqn Inference type (\ref{eq:onesample}) is
useful if only one sample is available or determining high
correlation regions within a group. Assuming the normality of
measurements $X$ and $Y$, the partial correlation $r=r_{Y_i,Y_j|X}$
can be transformed to be distributed as:
$$T = \frac{r\sqrt{n-2}}{\sqrt{1-r^2}} \sim t_{n-2},$$
the $t$ distribution with $n-2$ degrees of freedom. This test
statistic can be used for testing a one-sample inference type
(\ref{eq:onesample}).

The {\tt MATLAB} codes for computing the partial
correlation are given here. Let {\tt rho} be the sample partial
correlation  between cortical thickness ({\tt thick}) and
response time ({\tt time}) while removing the effect of age ({\tt
age}) and cortical area ({\tt area}) difference in a group at a
single vertex. For $n$ subjects in the group, all variables are row
vectors of size $1 \times n$. The {\tt MATLAB} codes for computing {\tt rho} is as follows: 
\begin{verbatim}
x=[age; area];
y=[thick; time];
a=cov([x;y]');
b=a(1:2,1:2)-a(1:2,3:4)*inv(a(3:4,3:4))*a(3:4,1:2);
rho=b(1,2)/sqrt(b(1,1)*b(2,2));
\end{verbatim}
Here {\tt x} and {\tt y} are $2 \times
n$ matrices, and the covariance matrix {\tt a} is the size $4 \times
4$.

\section{Surface-based Data Smoothing}
To increase the signal-to-noise ratio, we applied a surface based
smoothing method called {\em heat kernel smoothing} to the cortical
thickness measures. The implementation detail and its statistical
properties can be found in Chung et al. (2005). This is an improved
formulation over the previously developed {\em diffusion smoothing}
(Andrade et al., 2001; Chung et al., 2003; Cachia et al., 2003). In
Andrade {\em et al.} (2001) and Cachia {\em et al.} (2003),
smoothing is done by solving an isotropic heat equation via the
combination of the least squares estimation of the Laplace-Beltrami
operator and the finite difference method (FDM). In Chung {\em et
al.} (2003), the heat equation is solved using the finite element
method (FEM) and a similar FDM. The problem with these approaches to
data smoothing is the complexity of setting up the FEM and making
the FDM converge. Our heat kernel smoothing avoids all these
problems.

We assume the following linear model on thickness measure $Y$:
$$Y(p) = \theta(p) + \epsilon(p),$$
where $\theta(p)$ is the unknown mean thickness function and
$\epsilon(p)$ is a zero-mean random field, possibly a Gaussian white
noise process. Heat kernel smoothing of cortical thickness $Y$ is
then defined as the convolution: \bqn K_{\sigma} *Y
(p)=\int_{\partial \Omega}
K_{\sigma}(p,q)Y(q)\;dq\label{eq:conv},\eqn 
%\end{definition} 
where $K_{\sigma}$ is the heat kernel that generalizes the Gaussian kernel
in a Euclidan space to a curved manifold. The bandwidth $\sigma$
controls the amount of smoothing. Given the Laplace-Beltrami
operator $\Delta \psi = \lambda \psi$ on $\partial \Omega$, we can
order eigenvalues $0 = \lambda_0 \leq \lambda_1 \leq \lambda_2 \leq
\cdots$ and corresponding eigenfunctions $\psi_0,\psi_1,\cdots$. It
can be written in terms of basis function expansion:
$$K_{\sigma}*Y(p) = \sum_{j=0}^{\infty}  Y_j e^{-\lambda_j
\sigma} \psi_j(p),$$ 
where $$Y_j = \int_{\partial \Omega} \psi_j(q)Y(q)\;dq.$$
The {\em heat kernel estimator} of unknown functional signal $\theta(p)$ is then $$\widehat
\theta_{\sigma} (p) = K_{\sigma} * Y(p).$$ The heat kernel estimator
becomes unbiased as $\sigma \to 0$, i.e. $$\lim_{\sigma \to 0}
\mathbb{E} \widehat \theta_{\sigma}(p) = \theta(p).$$ As $\sigma$ gets
larger, the bias increases. However the total bias over all cortex
is always zero, i.e. $\int_{\partial \Omega} [\theta(p) - \mathbb{E}
\widehat \theta_{\sigma}(p)] \; dp = 0$. Further $$\lim_{\sigma \to
\infty} \widehat \theta_{\sigma} (p) = \frac{\int_{\partial \Omega} Y(q)
\;dq}{\int_{\partial \Omega} \;dq},$$ the sample mean over the whole
cortex $\partial \Omega$. Other properties of the heat kernel
smoothing can be found in Chung et al. (2005). The heat kernel
smoothing has been implemented in {\tt MATLAB} and it can be found
in the web {\tt
www.stat.wisc.edu/$\sim$mchung/softwares/hk/hk.html}. The approximate
relationship between the full width at half maximum (FWHM) and the
bandwidth is
$$\mbox{FWHM} = \sqrt{8\ln 2} \sigma.$$
In this study, the thickness measurements were smoothed with 30 mm
FWHM. This is the same amount of smoothing previously used in Chung
et al. (2005) for detecting cortical thinning in autism.

%\begin{figure}
%\centering
%\renewcommand{\baselinestretch}{1}
%\includegraphics[scale=0.15]{histogram.eps}
%\caption{\label{fig:taskscore} Z statistic.}
%\end{figure}

\section{Statistical Inference}

In our study, the main interest is testing the equality of
correlations between groups. So at each fixed point $p \in \partial
\Omega$, we are interested in testing \bqn H_0^B: \rho_1(p) =
\rho_2(p) \mbox{ vs. } H_1^B:\rho_1(p) \neq
\rho_2(p).\label{eq:twosample}\eqn For two sample inference type
(\ref{eq:twosample}), one approach is based on the {\em Fisher
transform} (Fisher, 1915; Hawkins, 1989; Bond and Richardson, 2004),
which shows the asymptotic normality:
$$ r_k \to \mbox{arctanh} (r_k) = \frac{1}{2}\ln \Big(\frac{1+r_k}{1-r_k}\Big) \sim
N\Big(\frac{1}{2}\ln
\Big(\frac{1+\rho_k}{1-\rho_k}\Big),\frac{1}{n_k-3} \Big).$$ The
transform can be viewed as a variance stabilizing normalization
process. Based on the Fisher transform, the test statistic under
$H_0^B$ is then given by: \bqn W(p) =\frac{\ln
\Big(\frac{1+r_1}{1-r_1}\cdot\frac{1-r_2}{1+r_2}
\Big)}{2\sqrt{\frac{1}{n_1-3}+\frac{1}{n_2-3}}} \sim N(0,1).
\label{eq:zfield}\eqn A slightly different formulation for testing
the equality of correlations can be found in Crawford et al. (2003).
We further normalized the field $W(p)$ with mean $\mu(p)=\mathbb{E}
W(p)$ and variance $S^2(p) = \mathbb{E} W^2(p) -\mu^2(p)$ by
$$Z(p) = \frac{  W(p) - \mu(p)}{S(p)}.$$
$\mu$ and $S^2$ are estimated from random permutations. We can take
the field $Z$ to be Gaussian with zero mean and unit variance. To
determine the null distribution of the test statistic, we permute
two samples across the groups. For $n_1$ subjects for group 1 and
$n_2$ subjects for group 2, we combine them together, do a random
permutation, and partition the result into two groups with the same
number of subjects. For this study, we generated 200 random
permutations out of $(n_1+n_2)!$ possible permutations. Then for
each permutation, we computed the statistic and based on the
empirical distribution of the statistic, we estimated $\mu$ and
$S^2$.

%\subsection{Nonparametric test}
%The nonparametric tests for correlation will be based on the {\em
%permutation test} first developed by R.A. Fisher
%(\cite{fisher.1966}) and used by many researchers in brain imaging
%({\bf put references}). Let $(y_{1j}^k,y_{2j}^k)$ be $n_k$
%measurements of variables $(Y_1,Y_2)$ for group $k$. Let $\tau$ be a
%permutation of indices $1,2,\cdots,n_2$. There are $n_2!$ different
%permutations. Then we permute measurements
%$(y_{1j}^k,y_{2\tau_i(j)}^k)$ and the corresponding sample partial
%correlation $r_k^*$. Then the null distribution of $r_k$ can be
%computed by
%$$P(r_k \leq h)  = \frac{\#\mbox{ of } r_k^*\leq h }{n_2!}.$$
%This distribution can be used for one sample inference $H_0^A$. For
%two sample inference $H_0^B$, we will use $R=\frac{1}{2}(r_1-r_2)$
%as the test statistic. To determine the null distribution of the
%test statistic, we permute two samples across the groups. For
%$n_1+n_2$ measurements $y_{21}^1, \cdots,y_{2n_1}^1,
%y_{21}^2,\cdots, y_{2n_2}^2$, we reindex them as $z_1,\cdots,
%z_{n_1+n_2}$. Then we permute them to $z_{\tau(1)},
%\cdots,z_{\tau(n_1+n_2)}$ and partition them into two groups
%$z_{\tau(1)}, \cdots,z_{\tau(n_1)}$ and $z_{\tau(n_1+1)},\cdots,
%z_{\tau(n_1+n_2)}$. Then we we have $(y_{1j}^1,z_{\tau(j)})$ for
%group 1 and $(y_{1j}^2,z_{\tau(n_1+j)})$ for group 2. The two groups
%produce the sample correlations denoted by $r_1^*$ and $r_2^*$. The
%distribution of $R$ is then given by
%$$P(R \leq h) = \frac{\#\mbox{ of } \frac{1}{2}(r_1^*-r_2^*)
%\leq h }{(n_1+n_2)!}.$$

Using $Z$ as the test statistic, we tested: \bq H_0: \rho_1(p) =
\rho_2(p) \mbox{ for all } p \in
\partial \Omega \mbox { vs.}\\
 H_1:\rho_1(p) \neq \rho_2(p) \mbox{ for some } p \in \partial
\Omega.\eq The null hypothesis $H_0$ is the intersection of
collection of hypotheses
$$H_0 = \bigcap_{p \in
\partial \Omega} H_0^B(p),$$ where $H_0^B$ is the null hypothesis given in
(\ref{eq:twosample}). The type I error $\alpha$ for testing one
sided test is then given by:
 \bq\alpha &=&
P\Big(\bigcup_{p \in
\partial \Omega} \{Z(p) >
h\}\Big)\\
&=& 1-  P\Big(\bigcap_{p \in \partial \Omega} Z(p) \leq h  \}\Big)\\
&=& 1 - P\Big(\sup_{p \in \partial \Omega} Z(p) \leq h \Big) \\
&=&
P\Big(\sup_{p \in
\partial \Omega} Z(p) > h \Big)\eq
for some $h$. The distribution of $\sup_{p \in \partial \Omega}
Z(p)$ is asymptotically given as: \bqn P(\sup_{p \in \partial
\Omega} Z(p)
> h) \approx \sum_{d=0}^2 \phi_d(\partial \Omega)\rho_d(h),
\label{eq:sup} \eqn where $\phi_d$ are the $d$-dimensional Minkowski
functionals of $\partial \Omega$ and $\rho_d$ are the
$d$-dimensional Euler characteristic (EC) density of correlation
field (Worsley et al., 1995). The Minkowski functionals are $\phi_0
=2,\phi_1 =0, \phi_2 = \mbox{area}(\partial \Omega)/2 = 49,616
 \mbox{mm}^2$, the half area of the template cortex $\partial
\Omega$. The EC densities are:
\bq \rho_0(h) &=& \int_{h}^{\infty}
\frac{1}{\sqrt{2\pi}}e^{-u^2/2}\;du \\
\rho_1(h)  & = &  \frac{(4\ln
2)^{1/2}}{2\pi \mbox{FWHM}} e^{-h^2/2} = \frac{1}{2\sqrt{2} \pi \sigma} e^{-h^2/2} \\
\rho_2(h) &=&   \frac{1}{(8\pi)^{3/2}\sigma^2}he^{-h^2/2} = \frac{4\ln
2}{(2\pi)^{3/2}\mbox{FWHM}^2}he^{-h^2/2}.\eq The resulting P-value
maps are found in Figure 1 and 2.

%\subsection{Nonparametric test}
%
%
%For two sample test, we are interested in testing
% \bqn H_0: \rho_1(p) = \rho_2(p) \mbox{ for all } p
%\in
%\partial \Omega \mbox{ vs. } H_1:\rho_1(p) \neq \rho_2(p) \mbox{ for some }
%p \in \partial \Omega \label{eq:multi2}.\eqn If we use
%$R(p)=\frac{1}{2}(r_1(p)-r_2(p))$ as the test statistic, the type I
%error $\alpha$ for testing (\ref{eq:multi2}) is given by $$\alpha =
%P\Big(\sup_{p \in \partial \Omega} R(p) > h \Big)$$ for some $h$.
%Unfortunately, we do not know the distribution of $\sup_{p \in
%\partial \Omega} R(p)$ so we can not apply the usual
%random field theory for correcting for the multiple comparisons.
%This gives a motivation for extending the permutation test presented
%in section 2.2.  For each permutation, we partition them into two
%groups as explained by section 2.2., and compute $\frac{1}{2}\sup_{p
%\in
%\partial \Omega} [r_1^*(p)-r_2^*(p)]$. Then we estimate the
%distribution of the $\sup_{p \in \partial \Omega}$ by
%
%$$P\Big(\sup_{p \in \partial \Omega} R(p) \leq h \Big)= \frac{\#
%\mbox{ of } \frac{1}{2}\sup_{p \in
%\partial \Omega} [r_1^*(p)-r_2^*(p)] \leq h}{(n_1+n_2)!}.$$

%\begin{figure}[t]
%\centering
%\renewcommand{\baselinestretch}{1}
%\includegraphics[scale=0.17]{hist.taskscore.eps}
%\caption{\label{fig:taskscore} Left: histogram of $\sup_{p \in
%\partial \Omega} R(p)$ based on 2400 permutations. Right: plots of $95\%$, $90\%$, $85\%$,
%$80\%$ (from top to bottom) upper percentiles over the number of
%permutations showing the convergence after approximately 2000
%permutations.}
%\end{figure}

\begin{figure}
\centering
\renewcommand{\baselinestretch}{1}
\includegraphics[width=0.98\linewidth]{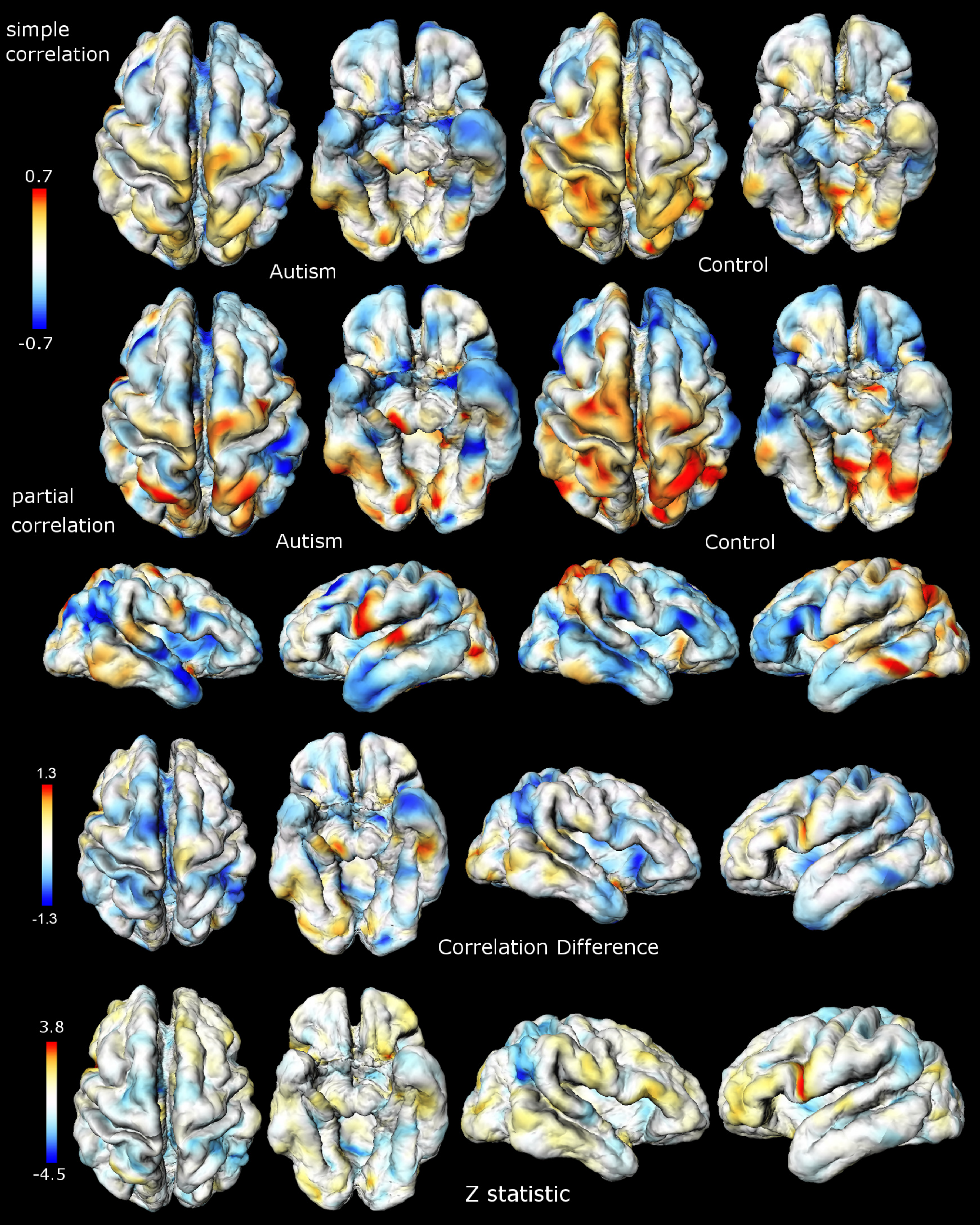}
\caption{\label{fig:taskscore} Map of facial emotion discrimination
task score correlated with thickness. The first raw is the simple
correlation. The second and third rows are the partial correlation.
The partial correlation tend to boost over all correlation values.
%increased the maximum correlation upto 0.9 from 0.7. 
The fourth row is the partial correlation difference
between the two groups (autism $-$ control). The last row shows the
final Z-statistic map showing statistically significant correlation
difference (P-value 0.03 for $z=3.8$, and 0.002 for $z=-4.5$).}
\end{figure}

\section{Application}

We applied our methodology to detect the regions of abnormal
brain-behavior correlates in autistic cortical regions.\\

 {\em Subjects.} 14 high functioning autistic (HFA) and 12
normal control (NC) subjects used in this study were screened to be
right-handed males. Age distributions for HFA and NC are $15.93 \pm
4.71$ and $17.08 \pm 2.78$ respectively. This is the same data set
used in previous studies Chung, et al. (2004), Chung, et al. (2005)
and Dalton, et al. (2005).\\

{\em Magnetic resonance images.} High resolution anatomical magnetic
resonance images (MRI) were obtained using a 3-Tesla GE SIGNA
(General Electric Medical Systems, Waukesha, WI) scanner with a
quadrature head RF coil. A three-dimensional, spoiled gradient-echo
(SPGR) pulse sequence was used to generate $T_1$-weighted images.
The imaging parameters were TR/TE $=21/8$ ms, flip angle $=
30^{\circ}$, 240 mm field of view, 256x192 in-plane acquisition
matrix (interpolated on the scanner to 256x256), and 128 axial
slices (1.2 mm thick) covering the whole brain.\\

{\em Cortical thickness.} Following image processing steps described
in Chung, et al. (2004) and Chung, et al. (2005) both the outer and
inner cortical surfaces were extracted for each subject via
deformable surface algorithm (MacDonald et al., 2000). Surface
normalization is performed by minimizing an objective function that
measures the global fit of two surfaces while maximizing the
smoothness of the deformation in such a way that the pattern of
gyral ridges are matched smoothly (Robbins, 2003; Chung et al.,
2005). Afterward cortical thickness was computed for each subject
(Chung et al., 2003; Chung et al., 2005). Heat kernel smoothing was
applied to the cortical thickness measures with a relatively large
30mm FWHM as described in a previous section.\\

{\em Facial emotion discrimination task.} The subjects were asked to
decide whether a picture of a human face was either emotional
(happiness, fear or anger) or neutral (showing no obvious emotion)
by pressing one of two buttons. The faces were black and white
photographs taken from the Karolinska Directed Emotional Faces set
(Lundqvist, et al., 1988; Dalton, et al., 2005). The task scores
(maximum 40) for HFA and NC are 27.14 $\pm$ 15.34 and 39.42 $\pm$
0.79 respectively, and the response time (ms) for HFA and NC are
$1329.8 \pm 206.7$ and $1110.9 \pm 182.3$ and respectively. A more
detailed description about the task can be found in Dalton et al.
(2005).\\

{\em Partial correlation maps.} The simple correlations between
cortical thickness and both task score and response time were
computed for each group and mapped onto the template cortex (Figure
1 and 2, first rows). The partial correlations were also computed
while removing the effect of age and global area difference. (Figure
1 and 2, second and third rows). Comparing the partial correlation
maps to the simple correlation maps, we see different patterns
indicating that it is necessary to account for the age and the area
terms for proper correlation analysis. The partial correlation
difference maps (autism $-$ control) show the regions of maximum
correlation difference (Figure 1 and 2, fourth rows). To access the
statistical significance of the correlation difference, the Fisher
transformation and the normalization steps were used resulting in
the Z-statistic maps (Figure 1 and 2, last rows).

Group difference between the autistic and control subjects were
identified using brain-behavior correlations of task score and
response time. Brain-behavior partial correlations of task score and
cortical thickness identified group differences in mainly two
cortical regions: right angular gyrus (area 39) and the left Broca's
area (area 44). The area 39 shows the positive correlation for the
control subjects while it shows the negative correlation for the
autistic subjects (corrected P-value 0.002, z-value -4.5). The area
44 shows the negative correlation for the control subjects while it
shows the positive correlation for the autistic subjects (corrected
P-value of 0.03, z-value 3.8).

%For example, increases in the cortical thickness of the left
%inferior frontal gyrus predicts enhanced task performance in autism
%but poorer task performance in controls (Figure 1 last row).
%Increased cortical thickness in both the left supramarginal gyrus
%and right inferior parietal lobule is associated with reduced task
%performance in autism whereas controls show reduced task performance
%with reducing cortical thickness (Fig.1, bottom row, right). Studies
%of lesions to the supramarginal gyrus show a reduced ability to
%judge emotion in the face (adolphs.pdf ; jin.pdf). The inferior
%parietal lobule functions in attention-related processing and dorsal
%visual stream processing, both of which have been implicated in
%autism (REF?).

For time-thickness correlation, we found more statistically
significant regions of difference that are consistent with previous
studies. In general, the spatial patterns of behavioral response
time and thickness correlation shows more negative correlation
(blue) than positive correlation (red) in the control subjects and
the pattern is opposite for the autistic subjects (Figure 2 second
row). Faster response time in the control subject are related to a
thicker right ventral and dorsal prefrontal cortex while they are
related to thinning in the same area in the autistic subject
(corrected P-value 0.001, z-value 4.6). We found correlation
difference in the left superior temporal gyrus and superior temporal
sulcus (corrected P-value 0.04, z-value -3.7) (Figure 2 last row).
The autistic subjects show an aberrant spatial pattern of
behavioral-thickness correlation in the right frontopolar region
(BA10), which shows a direct correlation between response time
duration and cortical thickness not seen in the control subjects. We
also found that slower responses in controls are related to a
thinner right inferior orbital frontal cortex but slower responses
in the autistic subject are independent of right orbital prefrontal
cortical thickness (corrected P-value 0.001, z-value 4.6).

\begin{figure}
\centering
\renewcommand{\baselinestretch}{1}
\includegraphics[width=1\linewidth]{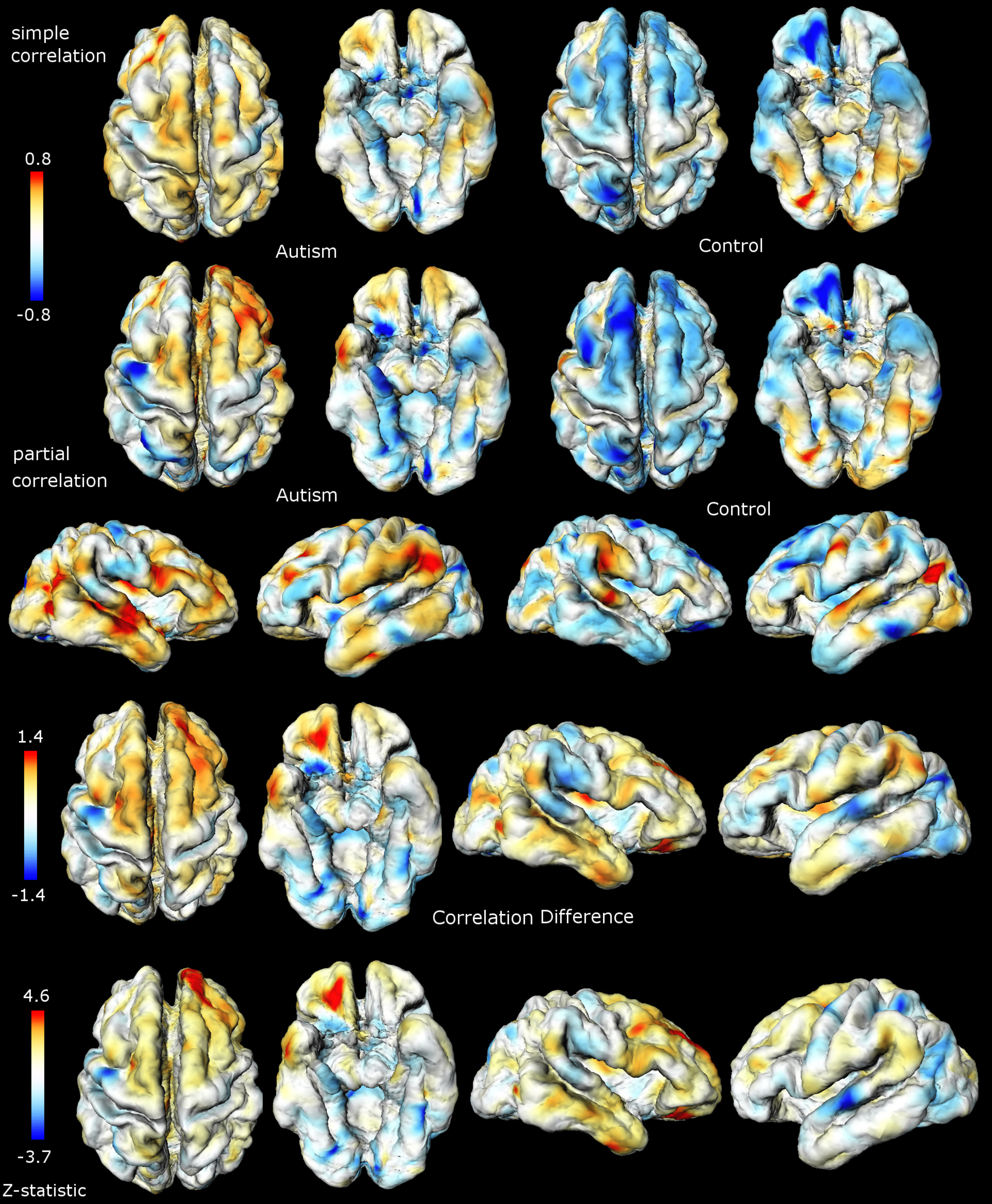}
\caption{\label{fig:taskscore} Map of response time correlated with
thickness. The first row shows the simple correlation. The second
and third rows are the partial correlation removing the effect of
age and cortical area differences. The fourth row shows the partial
correlation difference. We are interested in testing the
significance of this difference. The last row shows the final
Z-statistic map showing statistically significant correlation
difference (corrected P-value 0.04 for $z=-3.7$ and 0.001 for
$z=4.6$).}
\end{figure}

\section{Discussions}

In this study, group difference between the autistic and the control
subjects were identified using brain-behavior correlations between
cortical thickness and both task score and response time. The
partial correlation mapping strategy is shown to be an effective way
of visualizing and localizing the cortical regions of high
correlation while removing the effect of unwanted covariates such as
age, gender and global brain size differences. Our approach would be
a very useful analysis framework for many other types of future
brain-behavior correlate studies.

Our findings are consistent with previous functional and anatomical
studies. The score-thickness correlation difference found in the
left area 44 is interesting since this is the area shown to have
reduced bilateral connectivity in autism (Villalobos et al., 2005).
Since area 44 is thought to contain mirror neurons considered part
of the dorsal stream, altered brain-behavior correlations reflect
the influence of cortical thickness on perception-action function
(Rizzolatti et al., 2002; Villalobos et al., 2005).

The ventral prefrontal plays a role in the learning of tasks in
which subjects must learn to associate visual cues and responses
(Passingham et al., 2000; Passingham and  Toni, 2001). So our
finding of abnormal correlation between the response time and
thickness in the right ventral prefrontal cortex is not surprising.

Our previous study identified reduced cortical thickness in the
right inferior orbital prefrontal cortex, the left superior temporal
sulcus and the left occipito-temporal gyrus in the autistic subjects
relative to the control group (Chung et al., 2005). When paired with
results from our current study, areas in which cortical thickness is
reduced in autism predict differences in task response. Thicker
cortex in the left superior temporal gyrus and the superior temporal
sulcus predict faster response times in the control subjects,
whereas thicker cortex in the autistic subject are associated with
prolonged response times in these regions. This result may be
related to autistic dysfunction in the superior temporal gyrus and
superior temporal sulcus, regions known to be involved in social
processing (Baron-Cohen et al., 1999; Allison et al., 2000) and eye
gaze perception (Hooker et al., 2003). Slower responses in controls
are related to a thinner right inferior orbital frontal cortex but
slower responses in the autistic subject are independent of right
orbital prefrontal cortical thickness. This may suggest a floor
effect in which autistic cortical thickness is too thin to predict
changes in behavioral response time.

The general spatial patterns of behavioral response time-thickness
correlations distributed across the dorsal surface are positive in
the autistic subjects, whereas negative correlations are shown for
the control subjects in these regions. Autistics also show an
aberrant spatial pattern of behavioral-thickness correlation in the
right frontopolar region (BA10), which shows a direct correlation
between response time duration and cortical thickness not seen in
controls. One possible mechanism for these results is that increased
cortical thickness may produce alterations in intra cortical
connectivity resulting in a mis-allocation of cortical functional
resources. A recent study suggesting that alterations are noted
along the thickness of autistic cortex further complicates the
impact that alterations in autistic cortical anatomy may have on
behavior. Based on cellular studies, autistic subjects have an
increased number of smaller mini-columns, the basic functional unit
of cortex (Buxhoeveden and Casanova, 2002), that are less compact
relative to control subjects in prefrontal cortex and in temporal
regions. This anatomy may increase intracortical signalling, reduce
lateral inhibition, and cause terminal fields of subcortical
afferents to synapse on multiple mini-columns unintentionally
enhancing cortical noise in these regions (Casanova et al., 2002).
Our results add to this literature by identifying regions in which
cortical thickness alterations predict certain autistic behaviors.

%In summary, both positive and negative brain-behavior correlations
%are shown in regions with and without known group differences in
%cortical thickness between autistics and controls. These findings
%suggest that, over the course of development, the spatial
%distribution of cortical thickness alterations in autism influences
%the behavioral role that regions, both with and without cortical
%thickness alterations, may play in regulating emotional face
%processing accuracy and response times.

%\bibliography{hk.lncs}
%\bibliographystyle{plain}
%\bibliography{reference.april.2005}
\section*{References}

\begin{enumerate}

\item Andrade, A., Kherif, F., Mangin, J., Worsley, K.J., Paradis,
A., Simon, O., Dehaene, S., Le~Bihan, D., Poline, J-B. 2001.
Detection of FMRI activation using cortical surface mapping, Human
Brain Mapping 12, 79-93.

\item Allison, T., Puce, A., McCarthy, G. 2000. Social perception
from visual cues: role of the STS region.

\item Baron-Cohen,S., Ring, H.A., Wheelwright, S., Bullmore, E.T.,
Brammer, M.J., Simmons, A., Williams, S. C. R. 1999. Social
intelligence in the normal and autistic brain: an fMRI study.
European Journal of Neuroscience 11, 1891-1898.

\item Buxhoeveden, D.P., Casanova, M.F. 2002. The minicolumn
hypothesis in neuroscience. Brain 125, 935-951.

\item Bond, C.F., Richardson, K. 2004. Seeing the Fisher
Z-transformation. Psychometrika 60, 291-303.

\item Cachia, A. and Mangin, J.-F. and Rivi{\'e}re, D.,
Papadopoulos-Orfanos, D. and Kherif, F. and Bloch, I., R{\'e}gis,
J. 2003. A generic framework for parcellation of the cortical
surface into gyri using geodesic Vorono{\"\i} diagrams, Medical
Image Analysis 7,403-416.

\item Cao, J. and Worsley, K.J. 1998. The geometry of correlation
fields with an application to functional connectivity of the brain.
Annals of Applied Probability, 9, 1021-1057.

\item Casanova, M.F., Buxhoeveden, D.P., Switala, A.E., Roy, E.
2002.  Minicolumnar pathology in autism. Neurology, 58:428-432.

\item Chung, M.K., Worsley, K.J., Robbins,  S., Paus, P.,
Taylor, J., Giedd, J.N., Rapoport, J.L., Evans, A.C. 2003.
Deformation-based surface morphometry applied to gray matter
deformation, NeuroImage 18, 198-213.

\item Chung, M.K., Dalton, K.M., Alexander, A.L., Davidson, R.J.
 2004. Less white matter concentration in autism: 2D voxel-based
morphometry. NeuroImage 23,242-251.

\item Chung, M.K., Robbins,S., Dalton, K.M., Davidson,
Alexander, A.L., R.J., Evans, A.C. 2005. Cortical thickness analysis
in autism via heat kernel smoothing. NeuroImage 25, 1256-1265.

\item Crawford, J.R., Garthwaite, P.H., Howell, D.C., Venner, A.
2003. Intra-indiidual measures of association in neuropsychology:
inferential methods for comparaing a single case with a control or
normative sample. Journal of the International Neuropsychological
Society. 9, 989-1000.

\item Dalton, K.M., Nacewicz, B.M., Johnstone, T., Schaefer, H.S.,
Gernsbacher, M.A., Goldsmith, H.H., Alexander, A.L., Davidson, R.J.
2005. Gaze fixation and the neural circuitry of face processing in
autism. Nataure Neuroscience 8, 519-526.

\item Friston, K.J., Frith, C.D., Liddle, P.F., Frackowiak, R.S.J.
1993. Functional connectivity: The principal-component analysis of
large (PET) data sets. Journal of Cerebral Blood Flow and
Metabolism, 13:5-14

\item Friston, K.J., Frith, C.D., Fletcher, P., Liddle, P.F., Frackowiak,
R.S.J. 1996. Functional Topography: Multidimensional Scaling and
Functional Connectivity in the Brain Cerebral Cortex. Cerebral
Cortex. 6, 156-164.

\item Fisher, R.A. 1915. Frequency distribution of the values of the correlation coe±cient in
samples of an indefitely large population. Biometrika, 10, 507-521.

\item Hawkins, D.L. 1989. Using U statistics to derive the asymptotic distribution of Fisher's
Z statistic. The American Statistician, 43, 235-237.

\item Horwitz, B., Grady, C.L., Mentis, M.J., Pietrini, P.,
Ungerleider, L.G., Rapoport, S.I., Haxby, J.V. 1996. Brain
functional connectivity changes as task difficulty is altered.
NeuroImage, 3:S248.

\item Hooker, C.I., Paller, K.A., Gitelman, D.R., Parrish, T.B.,
Mesulam, M.-M., Reber, P.J. 2003. Brain networks for analyzing eye
gaze. Cognitive Brain Research. 17, 406-418.

\item Lundqist, D., Flykt, A., Ohman, A. 1998. Karolinska directed
emotional faces. Department of Neurosciences, Karolinska Hospital,
Stockholm, Sweden.

\item MacDonald, J.D., Kabani, N., Avis,  D.,
Evans, A.C. 2000. Automated 3-D Extraction of Inner and Outer
Surfaces of Cerebral
  Cortex from MRI. NeuroImage, {\bf 12}:340--356.

\item Passingham, R.E., Toni, I., Rushworth, M.F.S. 2000.
Specialisation within the prefrontal cortex: the ventral prefrontal
cortex and associative learning. Exp. Brain Res. 133, 103-113.

\item Passingham, R.E., Toni, I. 2001. Contrasting the dorsal and
venral visual systems: guidance of movement versus decision making.
NeuroImage 14:S124-31.

\item Rizzolatti, G., Fogassi, L., Gallese, V. 2002. Motor and
cognitive functions of the ventral premotor cortex. Cognitive
Neuroscience. 12, 149-154.

\item Robbins, S.M. 2003. Anatomical Standardization of the Human Brain
in Euclidean 3-Space and on the Cortical 2-Manifold. PhD thesis,
School of Computer Science, McGill University, Montreal, QC, Canada.

\item Thompson, P.M., Cannon, T.D., Narr, K.L., van Erp, T.,
Poutanen, V.P., Huttunen, M., Lonnqvist, J.,
Standertskjold-Nordenstam, C.G., Kaprio, J., Khaledy, M., Dail, R.,
Zoumalan, C.I., Toga, A.W. 2001. Genetic influences on brain
structure. Nature Neuroscience. 12, 1253-1258.

\item Villalobos, M.E., Mizuno, A., Dahl, B.C., Kemmotsu, N.,
Muller, R.-A. 2005. Reduced functionala connectivity between V1 and
inferior frontal cortex associated with visuomotor performance in
autism.

\item Worsley, K.J., Marrett, S., Neelin, P., Evans, A.C. 1995.
A unified statistical approach for determining significant signals
in location and scale space images of cerebral activation.
Quantification of brain function using PET, Eds. R. Myers,
V.J.Cunningham, D.L. Bailey and T. Jones , Academic Press,
San Diego, 327-333.

\item Worsley, K.J., Cao, J., Paus, T., Petrides, M., Evans, A.C.
1998. Applications of random field theory to functional
connectivity. Human Brain Mapping, 6, 364-367.

\item Worsley, K.J., Charil, A., Lerch, J., Evans, A.C. 2005.
Connectivity of anatomical and functional MRI data. The Proceeding
of the International Joint Conference on Neural Networks, Montreal,
Quebec, Canada.

\end{enumerate}

%\section*{Appendix}

\end{document}